# Efficient Operator State Migration for Cloud-Based Data Stream Management Systems

Jianbing Ding[1] Tom Z. J. Fu[2] Richard T. B. Ma[2,3]
Marianne Winslett[4] Yin Yang[5] Zhenjie Zhang[2] Hongyang Chao[6]

[1]School of Information Science and Technology, Sun Yat-sen University
[2]Advanced Digital Sciences Center, Illinois at Singapore Pte. Ltd.
[3]School of Computing, National University of Singapore
[4]Department of Computer Science, University of Illinois at Urbana-Champaign
[5]College of Science and Engineering, Hamad Bin Khalifa University
[6]School of Software, Sun Yat-sen University

dingsword@gmail.com, tom.fu@adsc.com.sg, tbma@comp.nus.edu.sg, winslett@illinois.edu, yyang@qf.org.qa, zhenjie@adsc.com.sg, isschhy@mail.sysu.edu.cn

## Abstract

A cloud-based data stream management system (DSMS) handles fast data by utilizing the massively parallel processing capabilities of the underlying platform. An important property of such a DSMS is *elasticity*, meaning that nodes can be dynamically added to or removed from an application to match the latter's workload, which may fluctuate in an unpredictable manner. For an application involving stateful operations such as aggregates, the addition / removal of nodes necessitates the *migration* of operator states. Although the importance of migration has been recognized in existing systems, two key problems remain largely neglected, namely *how to migrate* and *what to migrate*, i.e., the migration mechanism that reduces synchronization overhead and result delay during migration, and the selection of the optimal task assignment that minimizes migration costs. Consequently, migration in current systems typically incurs a high spike in result delay caused by expensive synchronization barriers and suboptimal task assignments. Motivated by this, we present the first comprehensive study on efficient operator states migration, and propose designs and algorithms that enable live, progressive, and optimized migrations. Extensive experiments using real data justify our performance claims.

## 1 Introduction

A data stream management system (DSMS) handles streaming data and answers continuous queries to users. Unlike traditional database systems, in a DSMS data is not available beforehand; meanwhile, key data properties such as arrival rates can fluctuate dynamically and unpredictably, meaning that the workload of a DSMS varies over time. Further, many streaming applications, such as video surveillance, have stringent response time constraints. To tackle these challenges, a popular solution is to base the DSMS on a cloud platform, which provides virtually infinite computational resources that can be elastically provisioned (e.g., by dynamically adding or removing nodes) to match the current workload. Nodes addition / removal is non-trivial when the application involves stateful operations such as aggregates, which necessitates moving contents of the states between nodes, a processed called *operator states migration* [11][12].

Figure 1a illustrates a simple execution plan for a word-count application containing two operators $Op_1$ and $Op_2$, which extracts and counts words from the input text stream, respectively. $Op_1$ is stateless whereas $Op_2$ is stateful, whose states are the word counters. Figure 1b zooms into $Op_2$ that is executed on three nodes $N_1$-$N_3$, each of which handles a horizontal partition of the input stream, and maintains the corresponding word counters. For ease of presentation, here the partitioning is based on the first letter of the word (e.g., $N_1$ deals with words whose first letter ranges from "a" to "i"). To achieve high processing speed, each node usually stores its corresponding operator states locally. Now suppose that a burst of inputs arrive, and in response the DSMS a new node $N_4$ is added to operator $Op_2$, shown in Figure 1c. To utilize $N_4$, $Op_2$ must divert some inputs to $N_4$, e.g., words starting with letters from "e" to "i", which were previous handled by $N_1$. However, $N_4$ cannot immediately process these inputs as it does not possess the corresponding counters; instead, these counters must be migrated from $N_1$ to $N_4$ before the latter starts working.

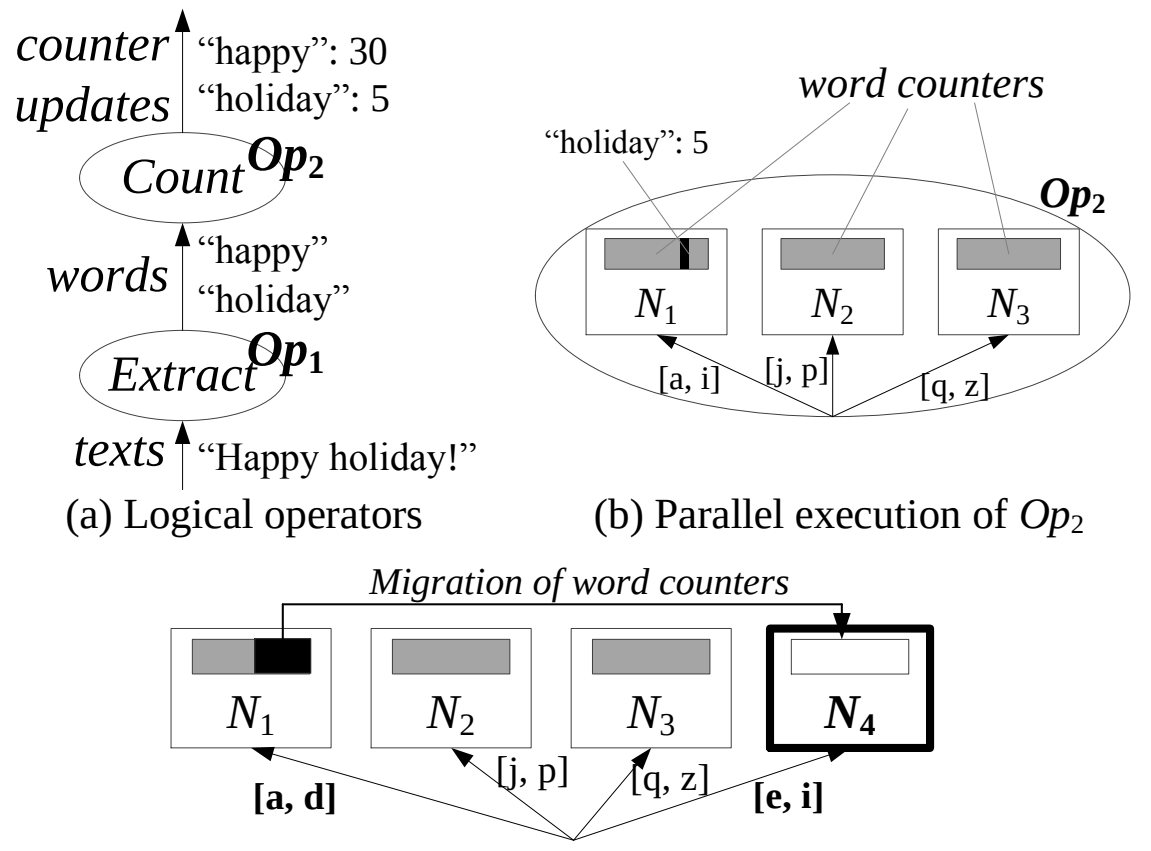

Figure 1. Example elastic execution in a massively parallel DSMS.

The concept of operator states migration is not new. However, as we review in Section 2, none of the existing system pays sufficient attention to optimizing migration efficiency. Notably, two fundamental problems have been largely neglected, namely, *how to migrate* and *what to migrate*. The first concerns how to reduce synchronization overhead and result delay during a migration. A naive approach involves 4 steps: suspending the entire operator, flushing all pipelines, moving operator states, and resuming the operator. Since this method suspends input processing, it may cause severe result delays, which is unacceptable in applications with real-time constraints. Worse, when the DSMS adds nodes in response to operator overload, the migration process exacerbates the problem by accumulating unprocessed new inputs. The second challenge we face is what operator states should be moved between nodes, which is determined by the new assignment of the input stream. In our example, the assignment shown in Figure 1c is just one of the numerous ways to re-partition the input stream among the 4 nodes after adding $N_4$. A commonly used strategy nowadays is *consistent hashing* (*CH*) [14], which minimizes the number of input partitions (called *tasks*) moved between nodes. For instance, a task in our example corresponds to input words with the same starting letter. As we elaborate in Section 2, CH is suboptimal when the tasks exhibit skewed workload and operator state sizes, leading to both load imbalance and high migration costs.

Motivated by this, we present the first comprehensive study on operator state migration, and propose efficient and effective solutions to tackle the above challenges. The proposed migration mechanism is capable of performing *live*, *progressive* migrations, meaning that the system continues to process new inputs while performing migration, and the impact of migration on result delay is small and controllable. Meanwhile, our task assignment algorithm computes the *optimal* assignment that minimizes migration costs while satisfying load balancing constraints. Moreover, an improved algorithm also considers the expected cost of future migrations, based on statistics collected from past workloads.

We have implemented the proposed methods based on Apache Storm [22], and evaluated them using a real dataset crawled from Twitter. The results confirm the performance advantages of these methods compared to existing ones.

# 2 Related Work

Earlier proposals of data stream management systems, e.g., Aurora [5], Borealis [3] and STREAM [4], mostly focus on centralized settings, or parallel settings with a fixed number of nodes. In other words, none of these systems supports elastic operator execution. One related problem, called *dynamic plan migration* [27], deals with the situation where the DSMS migrates to a completely different execution plan, e.g., a re-ordered join tree. This problem differs from ours in that the former moves states between operators, whereas we move states between nodes within an operator.

Recently, much interest has been shifted to cloud-based DSMSs, for which elasticity is a fundamental requirement. Existing cloud-based DSMSs can be classified into two categories: (i) operator-based DSMSs, e.g., Storm [22], S4 [17], etc., and (ii) mini-batch-based DSMSs, notably Spark Streaming [26]. The former resembles traditional DSMSs, whereas the latter shares features with cloud-based batch processing systems such as MapReduce [9] and Spark [25]. In particular, in a mini-batch-based DSMS, inputs are not processed immediately, but must wait until they form a batch of pre-defined size. Hence, such systems may not be ideal for applications with real-time constraints. Further, Spark Streaming in particular uses *immutable* states called RDDs, which makes it tricky to implement applications with mutable-state semantics such as counters. This paper focuses on operator-based DSMSs.

Among the existing systems, Storm [22] and S4 [17] are mature ones that have been widely used in practice. Their main difference is that S4 focuses on speed and provides no guarantee on result correctness, whereas Storm guarantees at-least-once semantics, i.e., each tuple is guaranteed to be processed, but possibly more than once. Neither of them, however, provides platform-level support for operator states, let alone migrations. Storm does support elasticity through its *rebalancing* mechanism, which allows node additions / removals. However, since Storm assumes stateless operators, all operator states are simply discarded by default during rebalancing.

Storm additionally contains an application-level add-on called Trident, which provides a rich set of features including support for operator states, transactions and exactly-once semantics. However, Trident incurs high overhead; in particular, in Trident a node does not store its operator states locally; instead, all states are maintained in an external transactional storage. This design renders migration trivial, at the expense of very high communication and computation costs as every update of an operator state requires a distributed commit. Samza [1] follows a similar design and incurs these costs as well.

An early version of our ongoing Resa project is overviewed in [21], which is a DSMS based on Storm featuring dynamic resource scheduling, operator state migration, and fault recovery. The early version implements the simple 4-step (suspend, flush, migrate and resume) migration describes in Section 1, which is costly. A Resa contributor independently built ChronoStream [24], which claims to have achieved migration with zero service disruption. Their migration implementation is bundled with the fault tolerance feature, which requires expensive I/O accesses to a local persistent storage. More importantly, as we explain in Section 3, their migration implementation can cause *incorrect results* due to synchronization issues. Finally, ChronoStream does not address the optimal target task assignment problem. SEEP [11] and StreamCloud [13] introduce the concept of operator states migration, but provide few details on how it is done in their system; neither system addresses the problems solved in this work.

Next we review task assignment solutions in a distributed DSMS, where a *task* corresponds to a partition of an operator's input stream. Consistent hashing (CH) [14] has been widely used for elastic task assignment, which guarantees, in a probabilistic sense, that (i) each node is assigned roughly the same number of tasks and (ii) during a migration, the number of re-assigned tasks is close to minimum. The problem of CH, however, is that not all tasks are equal, i.e., different tasks often exhibit different workload and operator state size due to data skewness. When this happens, CH obtains neither load balancing nor optimal migration cost (an example is provided in Appendix A). Gedik [12] designs a sophisticated hash function based on CH that achieves load balancing as well as small expected migration costs under the assumption that the size of operator states for each task is identical. The focus on [12] is hash function design rather than migration; in particular, it does not have any guarantee on the migration cost of a specific migration instance. Nevertheless, we compare with [12] in our experiments.

Finally, some concepts in this work, such as live migration, resemble those in traditional database migration. Notably, Albatross [10] migrates databases during a transaction, and Curino et al. [8] apply virtual machine migration methods to databases. The problems studied in these papers are fundamentally different from ours, and their methods are not applicable to our problems.

# 3 Migration Mechanism

Following common practice in existing systems, e.g., [11], [24], we assume that each node stores its corresponding operator states locally to obtain high processing efficiency; meanwhile, we assume that the workload of an operator is partitioned into tasks (e.g., words starting with the same letter), each with an *independent* operator states. Hence, during migration, it suffices to move tasks and their corresponding operator states between nodes.

## 3.1 Main Challenges

At first glance, migration appears easy: we simply move operator states according to the new task assignment through network transmissions; after that, we reroute the data flow of migrated tasks and release resources at their previous hosts. Meanwhile, Ref. [24] performs migration in parallel to task execution, which they claim to achieve "zero service disruption". In contrast, the suspend-flush-migrate-resume approach, e.g., used in [21], seems overly conservative and completely unnecessary.

The reality, however, is more complicated, due to several potential synchronization issues and network uncertainties. First of all, *a task cannot be migrated and executed at the same time*. Consider again the running example in Figure 1c. Suppose that the DSMS migrates the counters for all words starting with letter "e" to $N_4$, and at the same time $N_1$ continues to process such new inputs. Then, by the time $N_4$ receives these counters, some of them have already been obsolete, since $N_1$ has already updated them by processing new inputs. Consequently, if we simply redirect this partition of the input stream to $N_4$, the latter will produce incorrect results since it did not receive the inputs processed by $N_1$ during the migration. In other words, migration inevitably disrupts input processing, and the "zero service disruption" approach in [24] is incorrect. The challenge, then, is to minimize such disruptions through optimized migration mechanism design.

Second, *after a task is migrated, its old host may still receive its inputs*, due to a variety of reasons. For instance, in Figure 1c, after the DSMS finishes moving of counters for words starting with "e" to node $N_4$, and redirecting the corresponding slice of input stream to $N_4$, node $N_1$ can still encounter such tuples. A main reason for this is that distributed DSMSs commonly involve various buffers, e.g., to facilitate efficient network transmissions and inter-thread communications. Figure 2 illustrates inter-operator queues used in Storm [22]. Each *worker* (abstracted as a node in our terminology) can run multiple *executors* in parallel, each of which runs in a thread that handles multiple tasks. There are input and output queues on both worker and executor levels. After migrating a task (e.g., words starting with "e" from $N_1$ to $N_4$) and redirecting new inputs to the new host of the task ($N_4$), the old host ($N_1$) still needs to deal with inputs of this task in the input queues. Besides buffered tuples, there might be inputs to a task that are sent before its migration starts, and arrives at the old host after the migration. Additionally, if we do not place synchronization barriers at the beginning of a migration, different nodes may start the migration at different times. In this situation, a node may send an input based on the old task assignment to its host node before migration, leading to misrouted tuples.

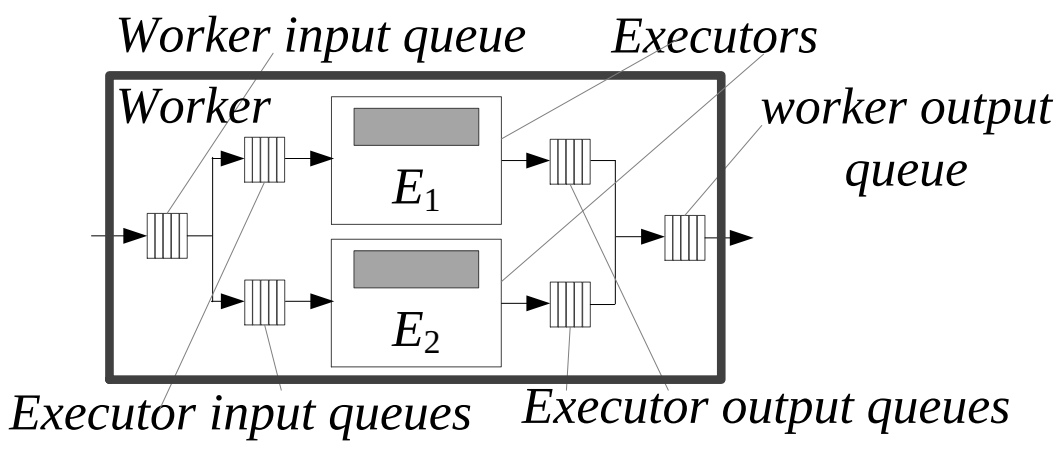

Figure 2. Tuple buffers in a Storm worker [22].

Third, *there are issues involved in the transmission of operator states*. One such issue is that the sender and receiver nodes of a task being migrated may be out of sync. In our running example, it is possible that when $N_4$ requests operator states from $N_1$, the latter may not have received the migration notice from the master node. Meanwhile, there may be networking issues, e.g., operator states can be lost during transmissions. Hence, we need a protocol to ensure that successful transmissions of operator states.

The above issues underline the difficulties to realizing migrations without enforcing expensive synchronization barriers such as suspending all inputs and flushing all buffered tuples. In the following, we present efficient and elegant solutions to handle these issues.

## 3.2 Proposed Migration Mechanism

The proposed migration mechanism performs live, progressive migrations without resorting to expensive synchronization barriers, and overcomes the challenges described in the previous subsection. Figure 3 shows the main components of the proposed migration mechanism, which include a *migration manager* (*MM*), a *retriever* of operator states from remote nodes, a *rerouter* for misrouted tuples, and an in-memory *file server* (e.g., Tachyon [15]) that serves operator states to remote retrievers. The MM, retriever and rerouters reside within the worker process (i.e., Java virtual machine), and the file server runs in a separate process to reduce system complexity, increase robustness, and reduce garbage collection costs [15]. The same file server can be shared by multiple workers running on the same physical machine.

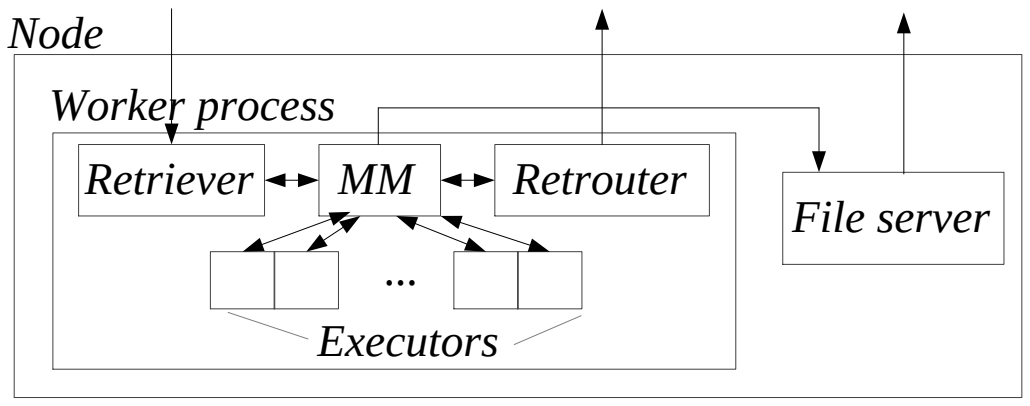

Figure 3. Key components of the proposed migration mechanism

When a migration is triggered, the master node (i.e., the *Nimbus* in Storm) asynchronously notifies all workers about the change in task assignment through Zookeeper [2]. Upon receiving such a notification, a worker updates its local task assignment, based on tuple routing is performed. Note that the new assignment takes effect at the beginning of the migration to minimize misrouted tuples. Then, the worker restarts the executors whose assigned tasks have changed. Here an executor is a thread within the worker process that can be restarted quickly without data loss, since all operator states and tuple buffers are stored on the process level. Meanwhile, the MM holds pointers to all operator states and tuple buffers of all executors, so that restarting an executor does not lead to its operator states or tuple buffers being deleted by the garbage collector. After an executor is restarted, it continues to execute tasks according to the new assignment. In particular, an executor may identify a misrouted tuple that belong to a task to be handled by another node, according to the new assignment. Such tuples include those in the input buffers as well as new ones received from the network. When this happens, the executor passes the misrouted tuples to the rerouter, which sends them to their respective host nodes according to the new task assignment.

A restarted executor may also find that the operator states of an incoming task are not yet ready. In this case, the task is suspended and all its new inputs must wait, until its states become available. As shown in Figure 4, we add a *wait buffer* for each such task in the node architecture. When the operator states of this task are ready, the executor flushes the buffer and removes the buffer, before processing new inputs. Clearly, it is important to limit number of waiting tuples, which is done through progressive migration, explained shortly.

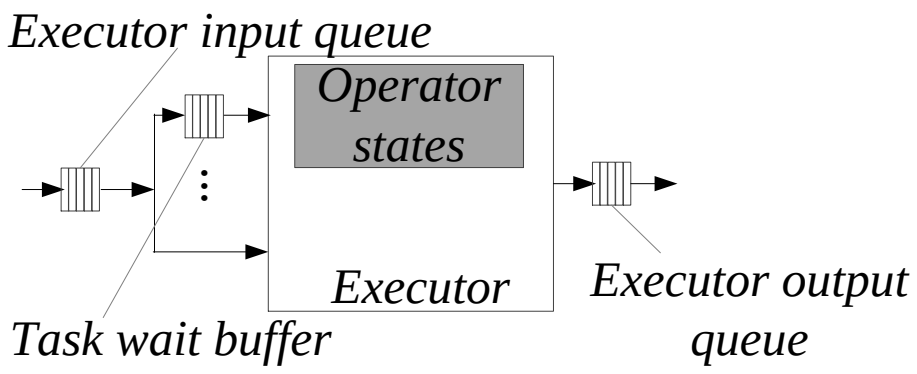

Figure 4. Task-lelvel input queues in the proposed migration mechanism.

In the meantime, the MM (i) sends the operator states of outgoing tasks to the file server, and deletes these operator states from the memory of the worker process, and (ii) instructs the retriever (which is a file server client) to download operator states of incoming tasks from the file servers of their corresponding nodes. When the download completes for a task, its corresponding executor starts processing its inputs accumulated at its task input queue.

The proposed mechanism clearly does not apply any synchronization barrier. It also achieves live migration since tasks that are not migrated continue to be executed during migration, and that migrated tasks start as soon as their operator states are delivered. This mechanism also overcomes the challenges described in the previous subsection. First, our mechanism never migrates and executes a task at the same time; in particular, when the MM moves the operator states of a task to the file server, the corresponding executor is also restarted and no longer processes the task. Second, the rerouters guarantee that all misrouted tuples are eventually sent to their correct destinations. Third, the file servers and their clients, i.e., the retrievers, use the file transfer protocol between them to ensure that all operator states are delivered correctly.

**Optimizations.** The transmission of operator states may cause *asymmetric network traffic*; for instance, one node with many outgoing tasks but few incoming ones may saturate its uplink bandwidth, while its downlink bandwidth remains mostly idle. We address this problem with the scheduling technique in [20], which achieves optimal total transmission time. The idea is to schedule transmissions in several phases, each of which saturates both the uplink and downlink traffic for every node.

In the mechanism described so far, new inputs of all tasks under migration must wait for their corresponding operator states, leading to increased result response time for these tuples. This can be a serious issue when the migration transmits operator states of large sizes. To alleviate the problem, we perform such migrations *progressively*: instead of updating the entire task assignment at once, we perform multiple mini-migrations, each of which migrates no more than a pre-defined number of tasks towards the new assignment. Accordingly, the increased result delay caused by migration becomes controllable.

# 4 Optimal Assignment Computation

This section focuses on the problem of finding the optimal task assignment that minimizes the costs for a single migration. Section 5 further tackles the problem that takes the costs of future migrations into consideration. Table 1 summarizes frequent notations in this section.

Table 1. List of common notations

| Symbol | Meaning |
|---|---|
| $n$ | Current number of nodes in an operator |
| $N_i$ | $i$-th node |
| $m$ | Number of tasks of an operator |
| $T_j$ | $j$-th task |
| $s_j$, $\lvert s_j \rvert$ | Operator state for task $T_j$ and its size |
| $I$=[$I.lb$, $I.ub$) | A task interval containing $T_{I.lb}$, $T_{I.lb+1}$,…, $T_{I.ub-1}$ |
| $I_i$, $I'_i$ | Task interval assigned to node $N_i$ before and after the migration, respectively |
| $\tau$ | Load balancing parameter (Definition 1) |
| $w_j$ | Amount of work for processing task $T_j$ |
| $W$ | Total amount of work for all tasks |

## 4.1 Problem Definition

We focus on finding the optimal task assignment for a stateful operator *Op*, currently executed on $n$ nodes. The input stream of *Op* is horizontally partitioned, and each node processes one such partition. Specifically, when a new input record $r$ arrives, *Op* first applies a partitioning function $f$ to $r$, and then routes $r$ to one of its nodes according to $f(r)$. Since such routing is performed on every input record, it is critical to minimize its overhead. Without loss of generality, we assume that the output of $f(r)$ is an integer in the range [1, $m$], where $m$ is a positive integer called the *number of tasks*. We use the term *task j*, (denoted as $T_j$, $1 \le j \le m$) to refer to the processing of all input records for which the partitioning function $f$ outputs $j$.

Following common cloud-based DSMS designs, e.g., in Storm [22], we assume that each node $N_i$ ($1 \le i \le n$) is assigned a continuous interval $I_i$=[$I_i.lb$, $I_i.ub$), $1 \le lb_i \le ub_i \le m$, called the *task interval* of $N_i$. The task intervals assigned to the $n$ nodes are mutually exclusive and collectively exhaustive with respect to all tasks in the operator. Each input $r$ is routed to node $N_i$, iff. $f(r) \in I_i$. This design enables fast routing, since the routing table, expressed as task intervals, is usually small enough to fit into CPU cache. If, for example, the routing is performed by an arbitrary mapping from tasks to nodes, the routing table will contain numerous entries, one for each task. Such a routing table probably does not fit into CPU cache, and, thus must reside in main memory, which is at least an order of magnitude slower than CPU cache [16].

A basic requirement in a parallel DSMS is *load balancing*, i.e., each node of an operator should have similar amount of work. Let $w_j$ denote the amount of work for task $T_j$, $W=\sum_{j=1}^{m} w_j$ be the total workload of the entire operator, and $W_i=\sum_{j \in I_i} w_j$ be the total workload for node $N_i$. We define the load balancing requirement as follows. Given $n$ nodes $N_1$-$N_n$ and a threshold $\tau$ ($\tau \ge 0$), a task assignment satisfies load balancing, iff. the workload $W_i$ for every node $N_i$ satisfies $W_i \le (1+\tau) \cdot W/n$. Intuitively, this definition requires that the workload of each node is not too high compared to the ideal case where every node has exactly the same amount work $W/n$. The smaller the threshold $\tau$, the tighter the requirement.

Next we define migration cost. Let $s_j$ ($1 \le j \le m$) be the operator state corresponding to task $T_j$, and $|s_j|$ be the size of $s_j$. We define the cost of a migration as the total size of operator states transmitted between nodes. The rationale is that the total transmission volume determines both the migration duration and the result delays of the affected inputs, as described in Section 3. The task assignment selection problem is defined as follows.

**Definition 4.1 (optimal single-step migration problem):** Given the current task assignment that involves $n$ nodes $N_1$-$N_n$ and their task intervals $I_1$-$I_n$, a target number of nodes $n'$ and a load balancing parameter $\tau$, find a target task assignment that minimizes the migration cost, while satisfying load balancing with threshold $\tau$.

## 4.2 Basic Solution

The computation of the optimal task assignment can be seen as two distinct steps: partitioning all tasks into $n'$ task intervals, and subsequently assigning these intervals to $n'$ nodes. Clearly, the load balancing constraint applies only to the partitioning step, and the migration cost is affected by both steps. Given a task partitioning, the assignment reduces to bipartite matching [23]. The challenge lies in the partitioning step, which has an exponential search space.

The proposed solution SSM (for single-step migration) simultaneously computes the best partitioning and the best assignment, using only $O(m \cdot n')$ space and $O(m^2 \cdot n')$ time. To simplify our notations, we assume that each node $N_i$ $1 \le i \le max(n, n')$ is assigned task intervals $I_i$ and $I'_i$ before and after the migration, respectively. Specifically, when $n' > n$, we assume that the new nodes are numbered $N_{n+1}$, $N_{n+2}$, …, $N_{n'}$; for each such node $N_i$ ($n < i \le n'$), we set $I_i$ to an empty interval. Similarly, when $n' < n$, we assign an empty interval $I'_i$ for each node $N_i$ removed after the migration. Meanwhile, without loss of generality, we assume that the $n$ nodes before migration are numbered $N_1$-$N_n$ in *ascending order* of their task intervals $I_1$-$I_n$ before migration. Since $I_1$-$I_n$ are disjoint and collectively exhaustive, we have $I_i.lb \le I_i.ub = I_{i+1}.lb$ for all $1 \le i < n$.

For each node $N_i$, $1 \le i \le n$, we define its *migration gain* as $g_i = \sum_{j \in I_i \cap I'_i} |s_j|$. Clearly, minimizing the migration cost is equivalent to maximizing the total gain, i.e., $\sum_i g_i$. An important observation is that our problem can be solved by dividing it into two independent sub-problems, recursively solving these sub-problems, and combining their solutions. We define a sub-problem as follows.

**Definition 4.2 (sub-problem of optimal single-step migration):** A sub-problem $P = \langle [\alpha, \beta), [\gamma, \delta), n_P \rangle$ ($1 \le \alpha < \beta \le m+1$, $1 \le \gamma < \delta \le n+1$, $1 \le n_P \le n'$) aims to partition tasks $T_\alpha$, $T_{\alpha+1}$, ..., $T_\beta$ into $n_P$ task intervals, and assign them to nodes $N_\gamma$, $N_{\gamma+1}$, …., $N_\delta$, such that the total gain of these nodes is maximized while satisfying load balancing.

Note that in the above definition, the number of nodes in sub-problem $P$, i.e., $\delta - \gamma$, is not necessarily equal to the target number of partitions $n_P$. When $\delta - \gamma > n_P$, there are $\delta - \gamma - n_P$ nodes that are not assigned a task interval in $P$, which we call *free nodes*. Similarly, when $\delta - \gamma < n_P$, there are $n_P - \delta + \gamma$ task intervals that are not assigned to a node, which we call free *task intervals*. A free node in one sub-problem can be either assigned a free interval in another sub-problem, or removed from the operator in case $n' > n$. The proposed algorithm SSM ensures that each free node always has zero gain. Hence, free nodes and intervals do not affect the optimal solutions of a sub-problem. According to the following lemma[1], $P$ can be solved by combining the solutions of its sub-problems.

**Lemma 4.1:** Given a sub-problem $P = \langle [\alpha, \beta), [\gamma, \delta), n_P \rangle$. Let *maxgain*($P$) be the total gain obtained by the optimal solution of $P$. We have:

$$maxgain(P) = max_{x,y,n_l}\big(maxgain(P_1) + maxgain(P_2)\big)$$

where $P_1 = \langle [\alpha, x), [\gamma, y+1), n_l \rangle$, $P_2 = \langle [x, \beta), [y+1, \delta), n_P - n_l \rangle$, $\alpha < x < \beta - 1$, $\gamma \le y < \delta - 2$, $1 \le n_l < n_P$.

[1] All proofs can be found in Appendix B.

The above lemma leads to a dynamic programming algorithm *Simple_SSM*, outlined in Figure 5. The main idea is to enumerate and solve each possible sub-problems $P$, and materialize the maximum gain $g[P]$ of $P$. In the base case that $n_P = 1$, solving $P$ is trivial, since $P$ assigns all tasks to a single node (line 6). Note that such an assignment can violate the load balancing requirement, in which case we set $g[P]$ to $-\infty$ (lines 4-5). When $n_P > 1$, Simple_SSM enumerates all possible combinations of $x$, $y$ and $n_l$ (line 9), and apply Lemma 4.1 to find the maximum gain of $P$ (lines 8-12). The values of $x$, $y$, $n_l$ that lead to this maximum gain are also stored, from which the optimal partitioning and assignment of tasks can be derived.

*Simple_SSM* ($m$, $n$, $n'$, $I_1$-$I_n$, $s_1$-$s_m$, $w_1$-$w_m$, $\tau$): output $I'_1$-$I'_{n'}$
// Inputs and outputs: refer to Table 1
1. For $n_P$=1 To $n'$
2. For each possible sub-problems $P = \langle [\alpha, \beta), [\gamma, \delta), n_P \rangle$
3. If $n_P$=1
4. If assigning all tasks in $[\alpha, \beta)$ violates load balancing
5. set $g[P]$ to $-\infty$
6. Else, solve $P$ and store its maximum gain in $g[P]$
7. Else
8. Initialize $g[P]$ to $-\infty$
9. For each possible combination of $x$, $y$, $n_l$
10. If $g[P_1]+g[P_2]>g[P]$
11. Update $g[P]$ to $g[P_1]+g[P_2]$
12. Store the values of $x$, $y$, $n_l$
13. Compute task intervals $I'_1$-$I'_{n'}$ that lead to the maximum gain of the original problem $\langle [1, m), [1, n), n' \rangle$, and return $I'_1$-$I'_{n'}$

Figure 5. Basic solution Simple_SSM for single-step migration.

## 4.3 Proposed Solution SSM

In Simple-SSM, there are $O(m^2 \cdot n^2 \cdot n')$ possible sub-problems, each requiring enumerating $O(m \cdot n \cdot n')$ combinations of $x$, $y$ and $n_l$, leading to a space complexity of $O(m^2 \cdot n^2 \cdot n')$ and time complexity $O(m^3 \cdot n^3 \cdot n'^2)$. The proposed algorithm SSM achieves significantly lower space and time complexities by exploiting a series of observations, as summarized in Table 2.

Table 2. Reduction of space and time complexities in SSM

| | **Space complexity** | **Time complexity** |
|---|---|---|
| Simple_SSM | $O(m^2 \cdot n^2 \cdot n')$ | $O(m^3 \cdot n^3 \cdot n'^2)$ |
| After Lemma 4.2 | $O(m \cdot n \cdot n')$ | $O(m^2 \cdot n^2 \cdot n'^2)$ |
| After Lemma 4.3 | $O(m \cdot n')$ | $O(m^2 \cdot n \cdot n'^2)$ |
| After Lemma 4.4 | $O(m \cdot n')$ | $O(m^2 \cdot n \cdot n')$ |
| After Lemma 4.5 (SSM) | $O(m \cdot n')$ | $O(m^2 \cdot n')$ |

The first observation is that in Lemma 4.1 node $N_y$ can be *any* node that has a non-zero gain in the optimal solution $P^*$ of $P$. When there are multiple such nodes in $P^*$, each would lead to a different division of $P$ into $P_1$ and $P_2$, and yet the same solution $P^*$ for $P$. SSM instead considers one such division: the one that minimizes $x$. Due to this requirement, $N_y$ must be the only node in $P_1$ with a non-zero gain, according to the following lemma.

**Lemma 4.2:** Given sub-problem $P$, there exists a combination of $x$, $y$, $n_l$ such that (i) *maxgain*($P$)=*maxgain*($P_1$)+*maxgain*($P_2$) and (ii) $N_y$ is the only node in $P_1$ with a non-zero gain in the optimal solution of $P_1$. The definitions of $P_1$, $P_2$ and the ranges of $x$, $y$, $n_l$ are the same as in Lemma 4.1.

Figure 6 shows algorithm *Solve_*$P_1$ that exploits the above observation to find the maximum gain of $P_1$ with *constant time and space*[2], assuming that there exists at least one feasible solution of $P_1$ that satisfies load balancing. The idea is to assign the longest task interval $I'_y$ to $N_y$ and return the gain of this assignment. Since the right boundary of $I'_y$ is fixed to $x$, maximizing its length also maximizes its gain. Meanwhile, doing so also minimizes the total amount of work for the remaining tasks, which ensures that there exists a way for partitioning these tasks into $n_l$–1 intervals under load balancing. Besides $I'_y$ and $N_y$, *Solve_*$P_1$ leaves the remaining tasks and nodes unassigned, which we deal with in a post-processing step.

---

*Solve_*$P_1(\alpha, x, \gamma, y, n_l)$: output $maxgain(P_1)$
// Inputs and output: refer to Lemma 4.1
// Assumption: a solution of $P_1$ exists that satisfies load balancing
1. Find the smallest $lb \geq \alpha$ such that task interval $I'_y=[lb, x)$ satisfies load balancing
2. Return the gain of assigning $I'_y$ to $N_y$

---

Figure 6. Algorithm for solving sub-problem $P_1$ in SSM

Since Solve_$P_1$ needs only constant time and space, there is no need to store its results; instead, we can solve $P_1$ on the fly every time we need its results. Hence, SSM only stores and reuses the results for sub-problem $P_2$. Observe that $P_2$ shares the same right boundaries $\beta$ and $\delta$ as its parent problem $P$. Hence, all such sub-problems have the same right boundaries $\beta$=$m$ and $\delta$=$n$. Thus, we reduce the space and time complexities to $O(m{\cdot}n{\cdot}n')$ space and $O(m^2{\cdot}n^2{\cdot}n'^2)$ time, respectively. Next we exploit another key observation: that most values of $y$ lead to *invalid* or *redundant* sub-problems, according to the following lemma.

**Lemma 4.3:** Given sub-problem $P$ and a combination of $x$, $y$, $n_l$ that satisfies both conditions in Lemma 4.2. (i) The task interval $I_y$ of node $N_y$ before migration satisfies $I_y.lb<x$; (ii) if $I_{y+1}.ub<x$, then $maxgain(P_2)=maxgain(P'_2=\langle[x, \beta), [y+2, \delta), n_P-n_l\rangle)$.

Given sub-problems $P$, $P_1$ and $P_2$ as in Lemma 4.1, According to the above lemma, $P_2$ is invalid when $I_y.lb>x$, and redundant when $I_{y+1}.ub \leq x$, whose optimal solution is identical to that of another sub-problem. In order for $P_2$ to be both valid and non-redundant, $I_y.lb<x$ and $I_{y+1}.ub \geq x$ must hold. Based on the assumption that nodes are numbered in ascending order of their task intervals before migration, $I_y.ub=I_{y+1}.lb$. Hence, either $I_y$ or $I_{y+1}$ contains $x$, meaning that given $x$, there are only two possible values of $y$ that lead to a valid and non-redundant $P_2$. This brings down the space complexity to $O(m{\cdot}n')$, and the time complexity to $O(m^2{\cdot}n{\cdot}n'^2)$.

Similar to the case of $y$, most values of $n_l$ also lead to redundant sub-problems, according to the following lemma.

**Lemma 4.4:** Given sub-problem $P$, there exists $x$ and $y$ such that the combination $x$, $y$, $n_{min}$ satisfies both conditions in Lemma 4.2, where $n_{min}$ is the minimum number of nodes for $P_1$ to have a feasible solution that satisfies load balancing given $x$ and $y$.

According to the above lemma, given $x$ and $y$, we can simply fix $n_l$ to $n_{min}$. This reduces the time complexity to $O(m^2{\cdot}n{\cdot}n')$. Finally, we cut another $n$ factor from the time complexity, using the following lemma.

**Lemma 4.5:** Given sub-problem $P$, suppose that the combination $x$, $y$, $n_l$ satisfies both conditions in Lemma 4.2. Then, either of the following two conditions holds (i) $x-1 \in I_y$, (ii) there does not exist another node $N_z$ in $P$ that leads to a higher gain for $P_1$, while satisfying $x \notin I_z$.

According to the above lemma, given $x$, we only need to enumerate two possible values of $y$, hence, the time complexity becomes $O(m^2{\cdot}n')$. Figure 7 shows the pseudo-code of SSM, which resembles Simple_SSM, with several key differences. First, according to Lemma 4.2, SSM fixes $\beta$=$m$ and $\delta$=$n$ (line 4) and invokes *Solve_*$P_1$ to solve sub-problem $P_1$ (line 17). Second, according to Lemma 4.3, given $\alpha$, SSM only enumerates two possible values for $\gamma$ (line 3); similarly, according to Lemma 4.5, given $x$, SSM only enumerates two possible $y$'s (line 16). Finally, according to Lemma 4.4, given $\alpha$ and $x$, SSM fixes $n_l$ to $n_{min}$ (line 17). The use of $n_{min}$ also guarantees that there is always a feasible solution for $P_1$, as required by Solve_$P_1$.

---

*SSM* $(m, n, n', I_1\text{-}I_n, s_1\text{-}s_m, w_1\text{-}w_m, \tau)$: $I'_1\text{-}I'_{n'}$
// Inputs and output: refer to Table 1
1. For $n_P$=1 To $n'$
2. For $\alpha$=1 To $m$
3. For each of the 2 possible values for $\gamma$ given $\alpha$
4. Set $P=\langle[\alpha, m), [\gamma, n), n_P\rangle$
5. If $n_P$=1, do lines 4-6 of Simple_SSM
6. Else
7. Initialize $g[P]$ to $-\infty$, $n_{min}$ to 1, $lb$ to 1, and $W_y$ to 0
8. For $x$=2 To $m$
9. Update $W_y$ to $W_y+w_{x-1}$
10. While $W_y > (1+\tau){\cdot}W/n'$
11. Increment $lb$ to $lb$+1
12. Update $W_y$ to $W_y-w_{lb}$
13. If $n_{min}$ nodes are insufficient for $P_1$, set $n_{min}$ to $n_{min}$+1
14. Set $P_1$, $P_2$ according to $x$, $y$, $n_{min}$
15. If $P_2$ is not redundant
16. For each of the 2 possible values for $y$ given $x$
17. Set $g$ to *Solve_*$P_1(\alpha, x, \gamma, y, n_{min})$
18. If $g+g[P_2]>g[P]$
19. Update $g[P]$ to $g+g[P_2]$
20. Store the values of $x$, $y$ and $n_{min}$
21. Compute task intervals $I'_1\text{-}I'_{n'}$ that lead to the maximum gain of the original problem $\langle[1, m), [1, n), n'\rangle$, and return $I'_1\text{-}I'_{n'}$

---

Figure 7. Proposed algorithm SSM for single-step migration

[2] As we explain later, both lines 1 and 2 in *Solve_*$P_1$ can be done in constant amortized time through careful bookkeeping.

We next clarify some important subtleties of the algorithm. The first is how algorithm $Solve_P_1$ finds the lower boundary of $I'_y$ in amortized constant time. Specifically, SSM maintains two variables $lb$ and $W_y$, initialized to 1 and 0 respectively (line 7), which satisfy $W_y = \sum_{lb}^{x-1} w_i$ . Whenever $x$ is incremented, we update $W_y$, and increase $lb$ to ensure that $W_y$ satisfies load balancing (lines 10-12). The value of $lb$ can then be used in $Solve_P_1$. Note that since $lb$ can only increase and its maximum value is $m$, updating $W_y$ and $lb$ take $O(m)$ time in total for all values of $x$; hence, each update of $W_y$ and $lb$ take amortized $O(1)$ time. Similarly, SSM incrementally updates $n_{min}$, the two possible values of $y$ given $x$, as well as the gain of $N_y$, such that each update of these values takes amortized $O(1)$ time. Finally, after obtaining the maximum gain of the entire problem, SSM performs a postprocessing step, which computes the optimal task partitioning and assignment that achieves this maximum gain using array $g$ as well as the stored values of $x$, $y$ and $n_{min}$ for each sub-problem, and assigns free nodes to free task intervals. We omit further details for brevity.

# 5 Multiple Migrations

The cost of single-step migration depends upon the current task assignment. Consequently, since the SSM algorithm only considers one migration, its resulting target assignment may lead to high costs for subsequent migrations. Hence, when for a sequence of migrations, applying SSM each time may lead to a sub-optimal total cost. Table 3 shows an example with 20 tasks. For simplicity, we make the following assumptions in this example: (i) the load balancing threshold $\tau$ is fixed to $\tau$ =0.4, (ii) the amount of work for every tasks is the same at all times, and (iii) the size of the operator state corresponding to each task is identical. Due to these assumptions, the total amount of work for a node, as well as the migration cost, can be simply expressed by the number of tasks involved, as shown in the table.

Table 3. Example migration sequence ($\tau$=0.4)

| Time | # of nodes | Single-Step | Alternative |
|---|---|---|---|
| $t_1$ | 2 | 13, 7 | 13, 7 |
| $t_2$ | 3 | 9, 9, 2 (cost = 4) | 8, 7, 5 (cost = 5) |
| $t_3$ | 4 | 6, 6, 2, 6 (cost = 6) | 6, 6, 4, 4 (cost = 4) |

Initially, at time $t_1$, there are 2 nodes $N_1$ and $N_2$, assigned 13 and 7 tasks respectively. Then, at $t_2$, we add one more node $N_3$ to the operator. According to the load balancing requirement, each node can handle at most 9 tasks. SSM assigns 9, 9 and 2 tasks to $N_1$-$N_3$, respectively, with a migration cost of 4 tasks, obtained by migrating two tasks from $N_1$ to $N_2$, and another two to from $N_1$ to $N_3$. After that, at time $t_3$, another node $N_4$ is added, and each node can now handle up to $20/4\times(1+0.4)=7$ tasks. SSM assigns 6, 6, 2 and 6 tasks to $N_1$-$N_4$, at a cost of 6 ($N_1$ and $N_2$ each migrating 3 tasks to $N_4$). The total cost for the two migrations is thus 4+6=10. Now consider an alternative solution shown in the rightmost column of Table 3, which incurs a sub-optimal cost at time $t_2$ (5 migrated tasks), but overall lower cost if we consider both $t_2$ and $t_3$ (5+4=9 migrated tasks). Hence, SSM does not always yield the lowest costs for a sequence of migrations.

## 5.1 Optimal Migration Sequence

We next study how to obtain optimal strategy for a sequence of migrations, assuming that we know the exact parameters of every one of them. This is often not possible in practice, and the reason we study this problem is to introduce concepts and algorithms that are reused in our main proposal in the next subsection. Specifically, similar to single-step migration, we are given load balancing factor $\tau$, $n$ nodes $N_1$-$N_n$ as well as their current task intervals $I_{0,1}$-$I_{0,n}$, and we are to do a sequence of $p$ migrations. Let $n_i$ be the target number of nodes at the $i$-th migration, and $I_{i,j}$ be the task interval of node $N_j$ after the $i$-th migration. Our goal is to find the optimal value for each such $I_{i,j}$, so that the total migration cost is minimized under load balancing.

First we observe that it is the *partitioning* of the tasks that matters to subsequent migrations, not the assignment, according to the following lemma.

**Lemma 5.1:** Given $n$ nodes before the migration, a sequence of $p$ migration parameters $n_1$-$n_p$, $\tau$, and two different initial node-to-task assignments $I_{0,1}$-$I_{0,n}$ and $I'_{0,1}$-$I'_{0,n}$. Suppose that there is a one-to-one mapping between $I_{0,1}$-$I_{0,n}$ and $I'_{0,1}$-$I'_{0,n}$. Then, the optimal solutions of migration sequence with respect to these two initial assignments have the same total cost.

Figure 8 shows the proposed solution OMS (optimal migration sequence). The main idea is to decompose the problem into two operations: finding the best strategy for the first migration, and solving the optimal migration sequence problem with the remaining $p$–1 migrations, starting with the task assignment after the first migration. OMS does this recursively, until there is only one migration left, at which point it applies SSM.

*OMS*($m$, $n$, $I_{0,1}$-$I_{0,n}$, $p$, $n_1$-$n_p$, $\tau_1$-$\tau_p$, $s_1$-$s_m$, $w_1$-$w_m$): output all $I_{i,j}$
// Inputs:
// $m$, $n$, $s_1$-$s_m$, $w_1$-$w_m$, $\tau$: same as in algorithm SSM
// $I_{0,1}$-$I_{0,n}$: initial task intervals for the $n$ nodes
// $p$, $n_1$-$n_p$, $\tau_1$-$\tau_p$: migration sequence parameters
// Outputs: task interval for each node after every migration

1. If $p$ = 1, return *SSM* ($m$, $n$, $n_1$, $I_{0,1}$-$I_{0,n}$, $s_1$-$s_m$, $w_1$-$w_m$, $\tau_1$)
2. For each partitioning of the $m$ tasks into $n_1$ task intervals that satisfy the load balancing requirement with parameter $\tau_1$
3. Compute the optimal task assignment from the initial state $I_{0,1}$-$I_{0,n}$ to the looping task partitioning, and the resulting task intervals $I_{1,1}$-$I_{0,n_1}$ after the migration
4. Call *OMS*($m$, $n_1$, $I_{1,1}$-$I_{0,n_1}$, $p$–1, $n_2$-$n_p$, $\tau_2$-$\tau_p$, $s_1$-$s_m$, $w_1$-$w_m$)
5. Compute the total migration cost by adding the migration cost of the first migration (line 3) and that of the remaining migrations (line 4).
6. Choose the partitioning that minimizes the total cost (line 5), set $I_{1,1}$-$I_{0,n_1}$ according to the result of line 3, and the rest of $I_{i,j}$'s according to the return value of OSM in lines 4.

Figure 8. Algorithm OSM for optimal sequence migration

Specifically, since the task assignment after the first migration affects subsequent migrations, OMS considers all such assignments. According to Lemma 5.1, it suffices to enumerate all possible partitionings of the $m$ tasks into $n_1$ intervals (line 2). For each such partitioning, OMS computes the optimal assignment from the initial task intervals to this task partitioning, yielding a concrete assignment $I_{1,1}$-$I_{0,n_1}$ after the first migration (line 3). This can be done with a standard matching algorithm, or a simplified version of SSM with the target partitioning fixed. After that, with $I_{1,1}$-$I_{0,n_1}$, OMS recursively solves the remaining assignments (line 4), until there is only one migration left, at which point OMS invokes SSM (line 1).

## 5.2 Optimal MTM-Aware Migration

OMS requires exact parameters of future migrations, which may not be practical. Fortunately, many applications do exhibit predictable patterns of their resource usage. For instance, on a social networking site, the workload depends on the users' activities, which correlates with time of the day. Such patterns are often probabilistic, e.g., at a future time, the system requires 3 nodes with a certain probability and 4 nodes with another probability. We capture such patterns with a *migration transition matrix* (*MTM*), a square matrix where each cell at the row $n$ and column $n'$ indicates the probability of migrating from $n$ to $n'$ nodes. Table 4 shows an example. An MTM can be computed using statistics of past server logs.

Table 4. Example migration transition matrix

| # of nodes | 2 | 3 | 4 |
|---|---|---|---|
| 2 | 0.3 | 0.6 | 0.1 |
| 3 | 0.3 | 0.4 | 0.3 |
| 4 | 0.1 | 0.5 | 0.4 |

Given a migration sequence, we calculate the probability of the sequence occurring using an MTM, assuming that each migration is independent of others. For example, according to Table 4, the probability of sequence 2, 3, 4 is 0.6×0.3=0.18, since the probabilities of migrating from 2 to 3 nodes and from 3 to 4 nodes are 0.6 and 0.3, respectively.

**Definition 5.2 (weighted sequence cost):** Given a sequence of $p$ costs $c_1$, $c_2$, …, $c_p$ and a discount factor $\gamma$ ($0\leq\gamma\leq1$), their weighted sequence cost is $\sum_{i=1}^{p}\gamma^{i-1}c_i$.

**Definition 5.3 (projected migration cost):** Given an MTM, a discount factor $\gamma$, and a task-to-node assignment, the latter's projected migration cost w.r.t. a random length-$p$ migration sequence is given by (i) enumerating every possible length-$p$ migration sequence, (ii) computing the optimal weighted sequence cost for each such sequence under load balancing and (iii) taking the weighted sum of their costs with their probabilities as weights.

Observe that when $\gamma$=1, the weighted sequence cost is simply the sum of migration costs for the entire sequence, and the projected migration cost is the expected optimal cost of a length-$p$ migration sequence. The intuition of parameter $\gamma$ is that the cost of a migrations happening in the far future is less important, due to accumulating uncertainty.

**Definition 5.4 (optimal MTM-aware migration problem):** Given an MTM, a discount factor $\gamma$, and parameters of a single-step migration, find an assignment that satisfies load balancing, while minimizing the sum of the (i) single-step migration cost and (ii) the projected migration cost for the task assignment after migration with respect to an infinitely long future migration sequence.

Next we solve the above problem. Similar to optimal migration sequence, it is the partitioning of the tasks, rather than the assignment, that affects to the cost of future migrations, according to the following lemma.

**Lemma 5.2:** Given initial assignments $I_{0,1}$-$I_{0,n}$ and $I'_{0,1}$-$I'_{0,n}$. Suppose that there is a one-to-one mapping between $I_{0,1}$-$I_{0,n}$ and $I'_{0,1}$-$I'_{0,n}$. Then, the projected migration costs of these two assignments are identical, regardless of the discount factor $\gamma$ and the length of the migration sequence $p$.

Based on the above lemma, the proposed solution works as follows: we enumerate all possible task partitionings with $n'$ nodes, and for each such partitioning we compute (i) its projected migration cost with respect to an infinite migration sequence, and (ii) the cost of the current migration, by matching the partitioning with the initial assignment $I_1$-$I_n$, as is done in OMS. Clearly, (i) is the expensive step. Fortunately, the projected migration cost of a partitioning can be pre-computed using Markov Decision Process [6][18], explained shortly. Let $n_{max}$ be the maximum number of nodes for the operator. In a pre-processing phase, we compute the projected migration cost for every partitioning of the $m$ tasks into up to $n_{max}$ task intervals using the MTM, and store and index these costs. Whenever we need to compute the optimal MTM-aware task assignment, we simply retrieve the pre-computed projected migration costs instead of re-computing them. This drastically cuts down the runtime cost of performing MTM-aware migration, rendering it practical.

Finally we clarify the pre-computation step. Figure 9 shows the proposed algorithm for computing the projected migration cost for one given task partitioning. The main idea is to follow the iterative solution for Markov Decision Process [6]. Specifically, initially all projected migration costs are set arbitrarily, e.g., to zero (line 1). Then, the algorithm iterates until the projected cost vector $C$ converges (lines 2-7). Inside each iteration, we re-compute the projected migration cost for each possible partitioning $P$, by performing one migration to any possible partitioning $P'$, and taking the weighted sum of the costs with weights determined by the MTM (lines 4-7). The correctness of the algorithm is guaranteed by the Markov Decision Process [6]. The computation of the cost for independent task partitionings can be done using a massively parallel processing system, such as MapReduce [9] and Spark [25].

```
PMC(m, n_max, s1-sm, w1-wm, τ, M): output vector C
// Inputs:
//     m, s1-sm, w1-wm, τ: same as in algorithm SSM
//     n_max: maximum number of nodes working for the operator
//     M: the migration transition matrix
// Outputs: a vector C containing the projected migration costs of //
//     every task partitioning w.r.t. an infinite migration sequence
Initialize C with arbitrary values
While C has not converged
    For each task partitionings P satisfying load balancing
        Initialize C[P] to 0
        For each task partitionings P' satisfying load balancing
            Compute the optimal migration cost c_{P,P'} from P to P'
            Update C[P] to C[P]+M[P,P'] ·(c_{P,P'}+γ·C[P'])
```

Figure 9. Algorithm for computing projected migration cost

# 6 Experiments

We implemented all proposed solutions based on Apache Storm [22] v0.9.1. Our code is publicly available at *https://github.com/ADSC-Cloud/resa*. The pre-computation module PMC (Figure 9) was implemented using Spark [25] v1.0.0. All experiments were run on a cluster of 9 servers, each equipped with an Intel Xeon quad-core 3.4GHz CPU and 24GB of RAM. We allocate one server to run as the master node, which hosts the Storm Nimbus and Apache Zookeeper [2]. In order to evaluate settings with a higher number of nodes, we run two nodes on each server, totaling 16 nodes. Each node is allocated 10 GB of RAM and two CPU cores.

The real dataset used in the experiments was crawled from Twitter. Specifically, the data contains 28,688,584 tweets from 2,168,939 users collected from October 2011 to November 2011. Over 90% of the data concentrate within a period of one month. From this dataset, we extract the number of required nodes at each timestamp, as follows. We partition time into intervals, each with the duration of one hour. Then, we count the number of tweets in each time interval, and allocate nodes proportional to the number of tweets. The number of nodes is normalized into the range of [8, 16]. If two adjacent time intervals have different number of nodes, we consider that a migration occurred between these two intervals; otherwise, no migration is done between these two intervals. For MTM-aware migration, we generate the MTM matrix as follows. For each possible number of nodes $n$, we count the number of times we migrate from $n$ nodes to $n'$ nodes, for all $n'$, using data from the previous month. These numbers are then normalized, as described in Section 5.2.

We tested two streaming applications in the experiments: word count and sliding window maximal frequent pattern mining [7] (referred to as "frequent pattern" in the following). The word count application is explained in Section 1. The frequent pattern application maintains a time-based sliding window of size $\omega$. Inside each window, we count the number of appearances for each combination of words (called "patterns"), and report frequent patterns, i.e., those whose appearance counts exceed a user-defined value. Figure 10 shows the operator topology of this application. Each tweet enters the application twice, once when it first arrives (counted as "+1") and once when it drops outside the sliding window (counted as "–1"). The Pattern Generator operator generates all patterns, i.e., word combinations, appearing in the tweet. The Detector operator maintains the counters for the appearances of these patterns, report frequent ones, and suppress those that are a subset of another frequent pattern. For instance, "Storm" is a subset of "Apache Storm" as the latter contains all the words of the former. Note that this is a rather complicated topology with a loop, which feeds the current frequent patterns back to the Detector itself, in order to suppress subsumed patterns.

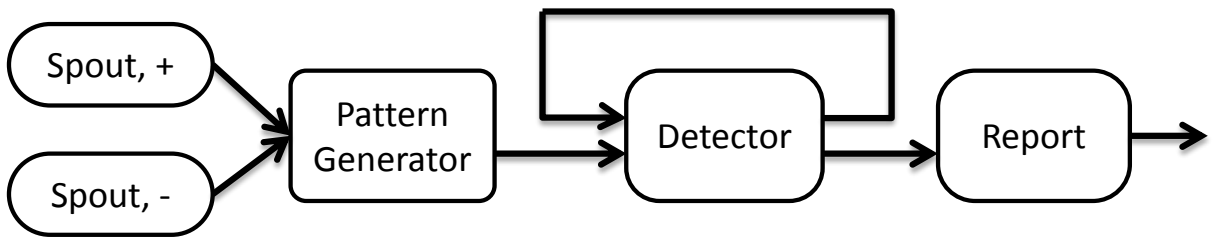

Figure 10. Operators for the frequent pattern application

We first evaluate our migration mechanism presented in Section 3. Due to lack of a suitable competitor, we modified Storm to allow operator states migration without data loss. Specifically, before a migration starts, all nodes write their operator states to a file server. Then, after the system re-starts, each node reads its assigned operator states from the file server. Figure 11 reports the average response time in each minute. The migration happens at the 7th minute, on which the number of workers changes from 10 to 8. The results show that our live migration reduces the average response time dramatically, by several orders of magnitude.

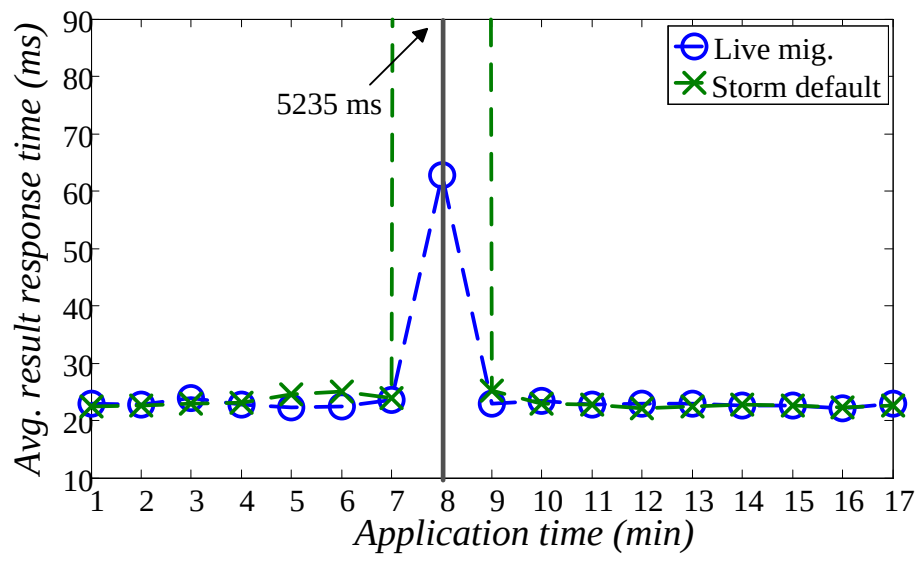

Figure 11. Evaluation of the proposed migration mechanism

Next we evaluate the task assignment computation algorithms. In each experiment, we vary one parameter, and fix all other parameters to their default values. Parameters evaluated in the experiments include (i) the load balancing threshold $\tau$, whose default value is 1.2, (ii) number of tasks $m$, whose default is 64, (iii) sliding window size $\omega$ in the frequent pattern application, whose default is 90 seconds, (iv) the discount factor $\gamma$ used in MTM-aware migrations, whose default is 0.8. Unless otherwise stated, in each experiment we run 100 consecutive migrations, and report the average performance (i.e., migration cost and running time) for one migration.

Figure 12 plots the migration cost as a function of the load balancing threshold $\tau$, comparing five migration strategies: optimal single-step, optimal MTM-aware, consistent hashing (CH) [14], Redist [12] and the default task scheduler of Storm [22]. Specifically, as described in Section 2, CH aims to have the same number of tasks on each node, and migrate the minimum number of tasks. The Storm scheduler is rather ad hoc, which simply allocates the same number of consecutive tasks to each node. Note that due to the random nature of CH, we run it for 100 times in each setting and report the average performance. The vertical axis is the total amount of transmissions during the migration, as a percentage of the total size of all operator states. For example, a number 30 on the vertical axis means that 30% of the all operator states are transmitted over the network during migration.

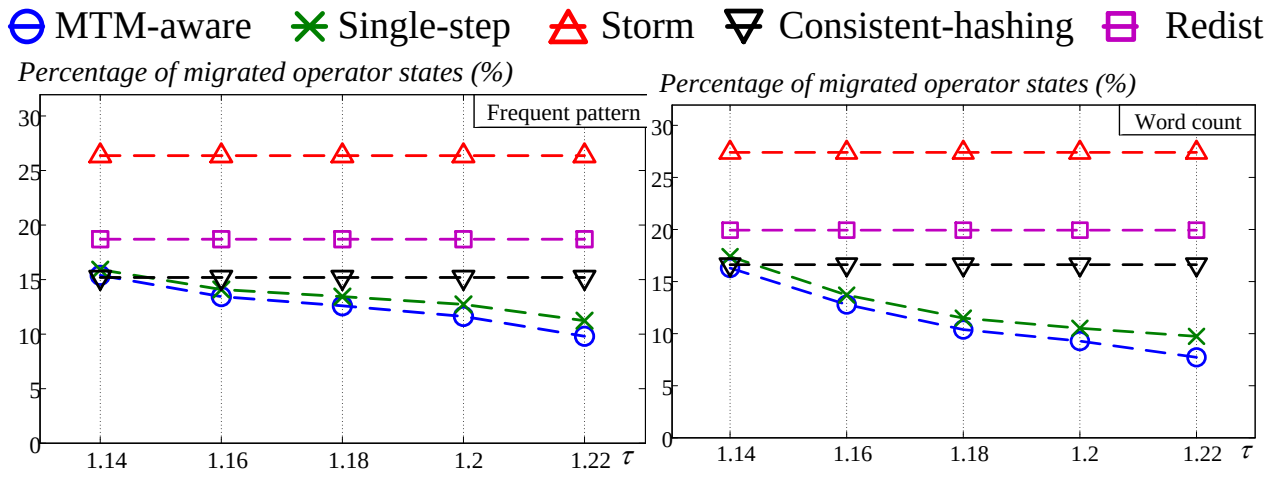

(a) Frequent Pattern (b) Word Count

Figure 12. Load balancing factor $\tau$ vs. migration cost

Comparing the 5 migration strategies, Storm clearly incurs a significantly higher migration cost compared to the two proposed methods. In particular, the cost of Storm is more than twice that of the proposed methods. The cost of CH is also consistently higher than the proposed solutions. Comparing single-step and MTM-aware migrations, the latter outperforms the former for all values of $\tau$. Their performance gap expands as $\tau$ increases, since a larger $\tau$ corresponds to a looser load balancing requirement, and, thus, a larger search space. Since single-step does not consider future migrations, its result is sub-optimal; hence, its performance gap with MTM-aware migration expands as the search space becomes larger. For similar reasons, the migration costs of the proposed methods decrease with increasing $\tau$, and the performance gap between them and the Storm and CH grows with $\tau$. Lastly, comparing the two applications, the migration cost decreases more rapidly with growing $\tau$ in word count than in frequent pattern mining. The reason is that word count is more susceptible to uneven data distributions; intuitively, a sudden burst of tweets containing a certain word can cause a sharp spike in its workload. As a result, when $\tau$ become larger, the search space expands rapidly as more skewed assignments become feasible. In contrast, frequent pattern mining is less sensitive to data distribution changes, since most patterns have low frequency, and are filtered early. Consequently, its migration cost is less sensitive to $\tau$.

Next we examine the running time of the proposed methods. For single-step migrations, we evaluate its total running time for computing the optimal task assignment, which is critical as it runs online. Figure 13 exhibits the running time of algorithm SSM with varying load balancing threshold $\tau$. As $\tau$ grows, the search space expands and the total running time of SSM grows. More importantly, the total running time of SSM never exceeds 2ms, which is negligible. Similarly, the running time of computing the optimal MTM-aware migration online is also negligible, and we omit its results for brevity.

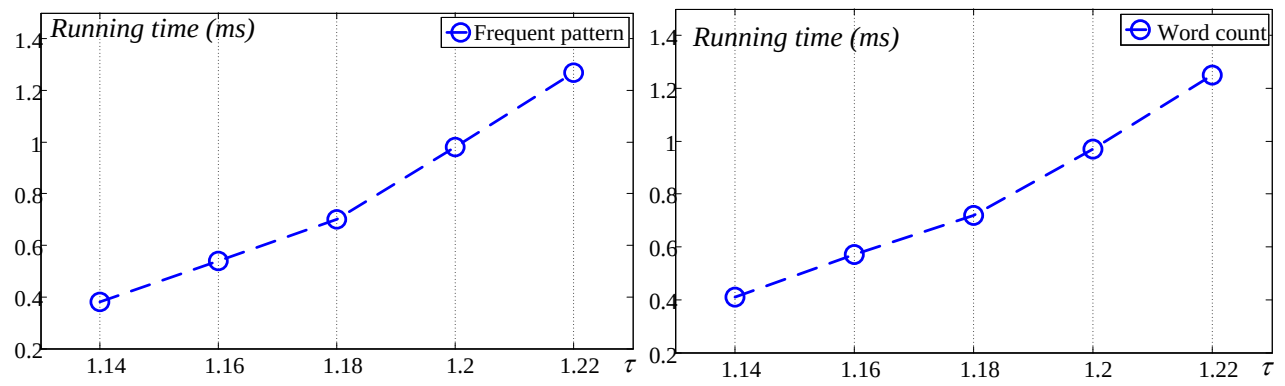

(a) Frequent Pattern (b) Word Count

Figure 13. Load balancing factor $\tau$ vs. running time of SSM

In addition to computing the optimal migration strategies, MTM-aware migration also involves a pre-computation module, using the PMC algorithm to compute the projected migration costs for each possible task partitioning. In our experiments we run PMC on Spark [25], using 64 executors (obtained by 8 machines × 4 cores × 2 hyperthreads per core). Figure 14 shows the pre-computation time with varying load balancing threshold $\tau$. Compared to online computation of the task assignment, the pre-computation takes a much longer time, ranging in hundreds of minutes. This indicates that MTM-aware migration successfully shifts most of the computations to the offline module. Considering that MTM-aware migration achieves consistently lower migration costs, the offline pre-computation time is a price worth paying for.

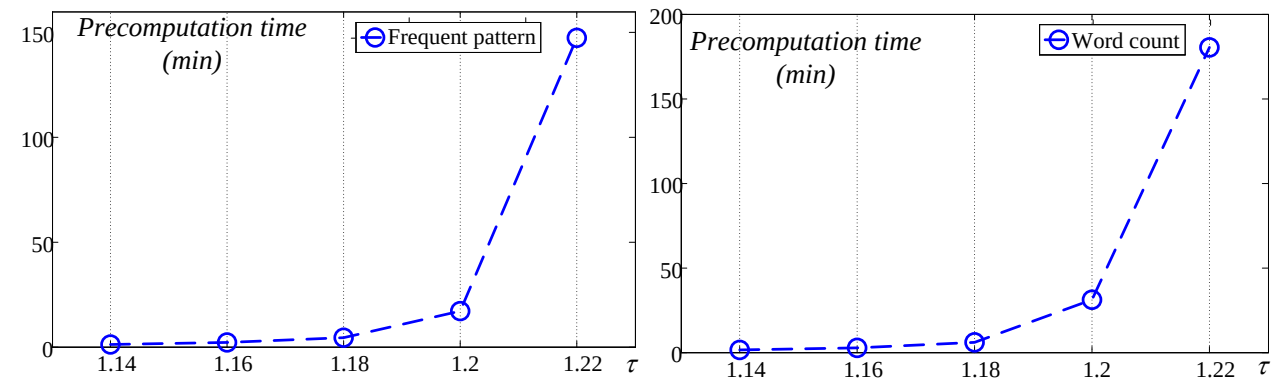

(a) Frequent Pattern (b) Word Count

Figure 14. Load balancing factor $\tau$ vs. running time of PMC

We next evaluate the impacts of other important parameters. Figure 15 illustrates the effect of the total number of tasks $m$. Recall from Section 4.1 that $m$ is determined by the partitioning function of the operator, which is user-configurable. Clearly, when $m$ grows, the running time of SSM increases quadratically, since its complexity is $O(m^2 \cdot n \cdot n')$ as described in Section 4. Meanwhile, similar to the case of growing $\tau$, a larger $m$ also increases the search space, as tasks are now in a finer granularity. As a result, the migration cost decreases with growing $m$, as SSM obtains higher quality migration strategies in a larger search space. The results for the frequent pattern application, as well as those for MTM-aware migration lead to similar conclusions, and are omitted for brevity.

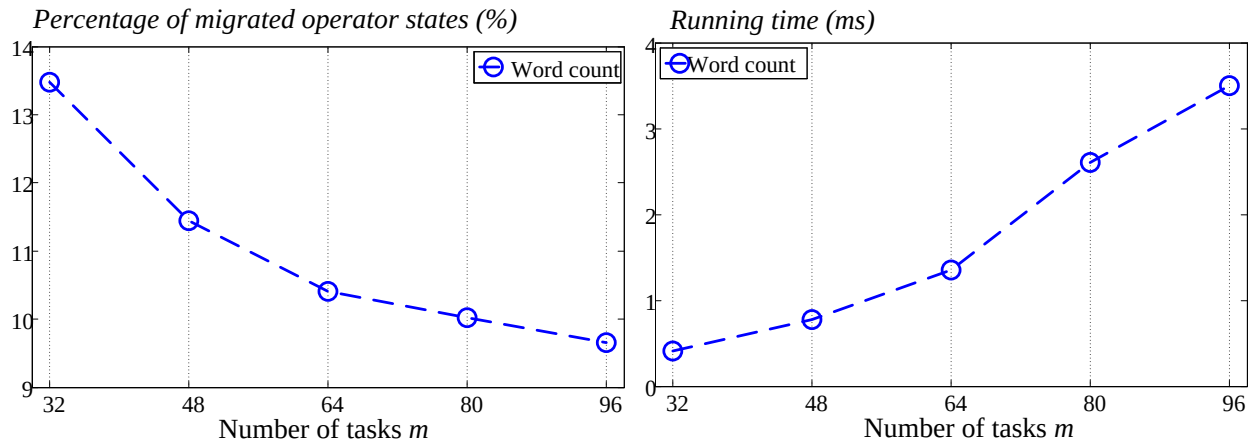

(a) Migration cost (b) Running time

Figure 15. Effects of number of tasks on single-step migration

Next we evaluate the actual result response time for different migration strategies. Specifically, we measure the delay of each result during the 2 minutes after the migration starts, and report the average delay. After that (i.e., 2 min after migration), the response time of all methods reaches a comparable steady state that is not affected by the migration process. To obtain accurate measurements on response time, we manually ensured that no other processes were running on the same machines and the network traffic was light during the experiments. Figure 16 shows the response time of the proposed migration methods as well as CH, using the frequent pattern application with varying window sizes. The response time of all methods grows with the window size, as a larger window leads a higher workload. Clearly, the proposed methods outperform CH by large margins. Comparing Figure 16 with previous experimental results, the performance gap between single-step and MTM-aware migration strategies is more pronounced in terms of response time than migration cost. The response time for Storm is at least 5000 milliseconds, even for the smallest window size. We thus exclude it from the diagram in order not to distort the scale.

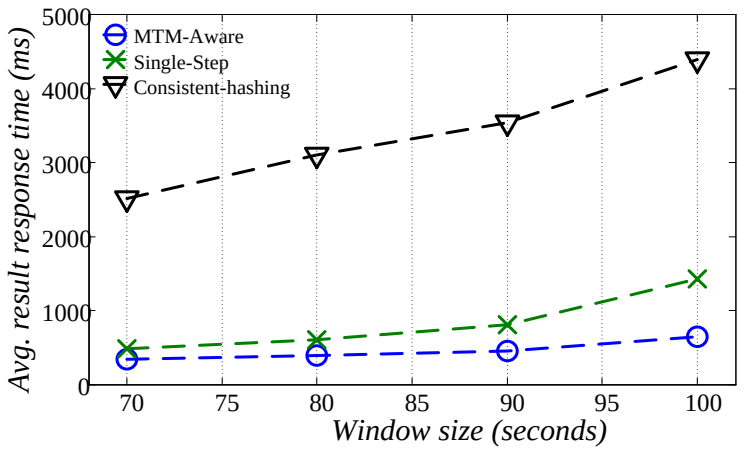

(a) Mean response times

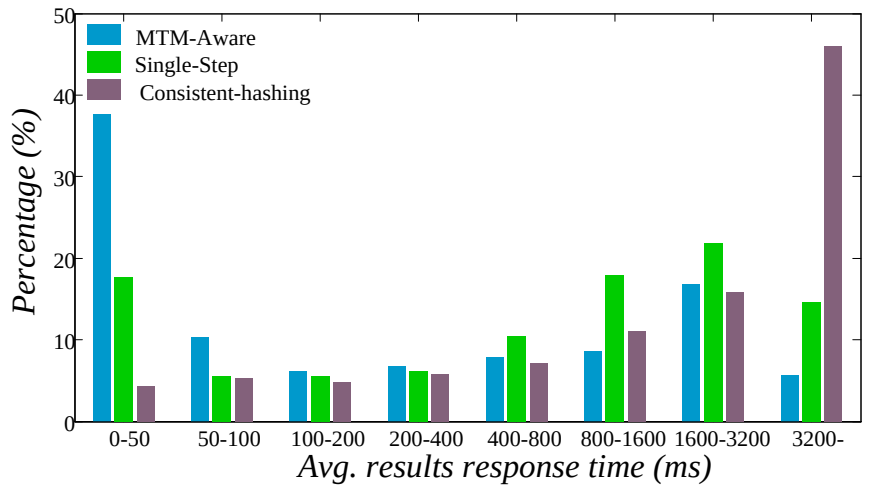

(b) Distribution of response times (window size: 90 seconds)

Figure 16. Window size vs. response time (FP)

Figure 17 shows the results of MTM-aware migration, with varying discount factor $\gamma$. According to the discussions in Section 5.2, a larger $\gamma$ places more emphasis on future migrations, and vice versa; as an extreme case, when $\gamma$=0, MTM-aware migration reduces to single-step migration. The experimental results confirm this. In particular, when $\gamma$ grows, the migration cost decreases, because the method now considers faraway future migrations. However, the pre-computation cost also becomes higher. Hence, by tuning $\gamma$, the user controls the tradeoff between migration cost and pre-computation overhead.

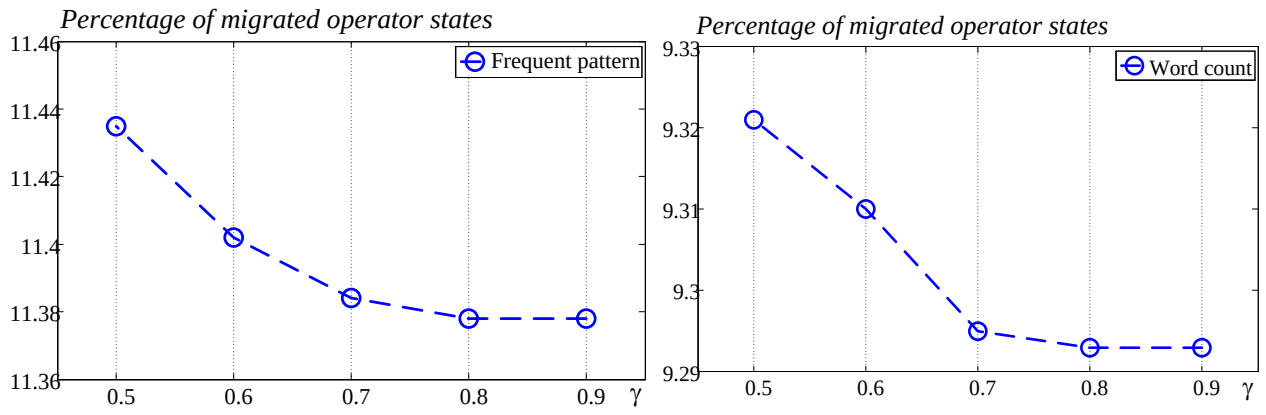

(a) Mig. Cost (Frequent Pattern) (b) Mig. Cost (Word Count)

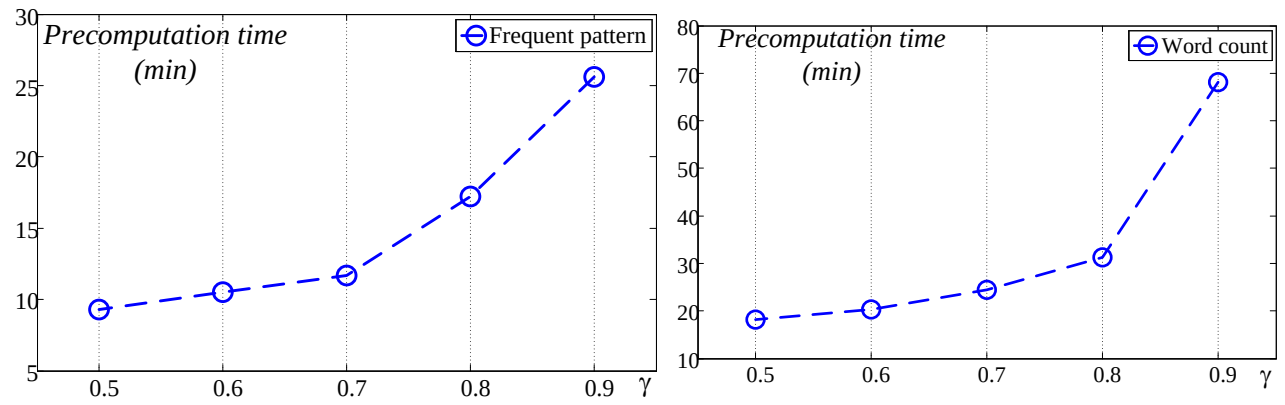

(c) Comp. Cost (Freq. Pattern) (d) Comp. Cost (Word Count)

Figure 17. Effect of Discount factor $\gamma$ on MTM

Summarizing the experimental results, the migration strategies obtained through the proposed methods achieve significant savings in terms of transmission overhead, compared to consistent hashing and the Storm default scheduler. Optimal MTM-aware migration yields higher quality migration strategies, at the cost of a pre-computation phase. This quality-cost tradeoff can be controlled through a tunable parameter $\gamma$. The computation overhead for both proposed task assignment computation techniques is small. Further, our design and implementation of the migration process achieves dramatic savings in terms of result response time. Overall, we believe that our migration strategies and system designs may be key enablers for a more elastic data stream management system.

# 7 Conclusion

We have systematically investigated the problem of efficient operator state migration in a cloud-based DSMS, and proposed a novel migration mechanism based on in Storm [22], as well as algorithms for finding the optimal task assignment for the migration. Extensive experiments with real data confirmed the effectiveness and efficiency of the proposed solutions. Regarding future work, an interesting direction is migrations that occur to multiple operators simultaneously, e.g., when the system shifts one node from one operator to another. In this case, interactions between nodes working for different operators may lead to opportunities for further optimization. Another direction is to apply our migration techniques to fault recovery, which involves periodically checkpointing each operator's states, and restoring them when a fault happens. An optimal strategy should strike a balance between recovery time and checkpointing overhead.

# 8 References

[1] Apache Samsa. *http://samza.apache.org* .

[2] Apache Zookeeper. *http://zookeeper.apache.org* .

[3] Abadi, D. J., Ahmad, Y., Balazinska, M., Çetintemel, U., Cherniack, M., Hwang, J.-H., Lindner, W., Maskey, A., Rasin, A., Ryvkina, E., Tatbul, N., Xing, Y., Zdonik, S. B. The Design of the Borealis Stream Processing Engine. *CIDR*, 2005.

[4] Arasu, A., Babcock, B., Babu, S., Datar, M., Ito, K., Nishizawa, I., Rosenstein, J., Widom, J. STREAM: the Stanford Stream Data Manager. *SIGMOD*, 2003.

[5] Abadi, D. J., Carney, D., Çetintemel, U., Cherniack, M., Convey, C., Lee, S., Stonebraker, M., Tatbul, N., Zdonik, S. B. Aurora: A New Model and Architecture for Data Stream Management. *VLDB J.*, 12(2): 120-139, 2003.

[6] Bellman, R. A Markovian Decision Process. *Journal of Mathematics and Mechanics*, 6, 1957.

[7] Burdick, D., Calimlim, M., Gehrke, J. Mafia: A Maximal Frequent Itemset Algorithm for Transactional Databases. *ICDE*, 2001.

[8] Curino, C., Jones, E. P. C., Madden, S., Balakrishnan, H. Workload-Aware Database Monitoring and Consolidation. *SIGMOD*, 2011.

[9] Dean, J., Ghemawat, S. MapReduce: Simplified Data Processing on Large Clusters. *OSDI*, 2004.

[10] Das, S., Nishimura, S., Agrawal, D., Abaddi, A. E. Albatross: Lightweight Elasticity in Shared Storage Databases for the Cloud using Live Data Migration. *PVLDB*, 4(8): 494-505, 2011.

[11] Fernandez, R. C., Migliavacca, M., Kalyvianaki, E., Pietzuch, P. Integrating Scale Out and Fault Tolerance in Stream Processing using Operator State Management. *SIGMOD*, 2013.

[12] Gedik, B. Partitioning Functions for Stateful Data Parallelism in Stream Processing. *VLDB J.*, 23(4): 517-539, 2014.

[13] Gulisano, V., Jiménez-Peris, R., Patiño-Martínez, M., Soriente, C., Valduriez, P. StreamCloud: an Elastic and Scalable Data Stream System. *IEEE TPDS*, 23(12): 2351-2365, 2012.

[14] Karger, D., Lehman, E., Leighton, T., Panigrahy, R., Levine, M., Lewin, D. Consistent Hashing and Random Trees: Distributed Caching Protocols for Relieving Hot Spots on the World Wide Web. *STOC*, 1997.

[15] Li, H., Ghodsi, A., Zaharia, M., Baldschwieler, E., Shenker, S., Stoica, I. Tachyon: Memory Throughput I/O for Cluster Computing Frameworks. *LADIS*, 2013.

[16] Manegold, S., Boncz, P., Kersten, M. Optimizing Main-Memory Join on Modern Hardware. *IEEE TKDE*, 14(4):709-730, 2002.

[17] Neumeyer, L., Robbins, B., Nair, A., Kesari, A. S4: Distributed Stream Computing Platform. *ICDM KDCloud*, 2010.

[18] Puterman, M. L., *Markov Decision Processes: Discrete Stochastic Dynamic Programming*, John Wiley & Sons, 2009.

[19] Qian, Z., He, Y., Su, C., Wu, Z., Zhu, H., Zhang, T., Zhou, L., Yu, Y., Zhang, Z. TimeStream: Reliable Stream Computation in the Cloud. *EuroSys*, 2013.

[20] Rödiger, W., Mühlbauer, T., Unterbrunner, P., Reiser, A., Kemper, A., Neumann, T. Locality-Sensitive Operators for Parallel Main-Memory Database Clusters. *ICDE*, 2014.

[21] Tan, T, Ma, R., Winslett, M., Yang, Y., Yong, Y., Zhang, Z. Resa: Realtime Elastic Streaming Analytics in the Cloud. *SIGMOD*, 2013, poster.

[22] Toshniwal, A., Taneja, S., Shukla, A., Ramasamy, K., Patel, J. M., Kulkarni, S., Jackson, J., Gade, K., Fu, M., Donham, et al. Storm@Twitter. *SIGMOD*, 2014.

[23] U, L. H., Mouratidis, K., Yiu, M. L., Mamoulis, N. Optimal Matching between Spatial Datasets under Capacity Constraints. *ACM TODS*, 35(2), 2010.

[24] Wu, Y., Tan, K. ChronoStream: Elastic Stateful Stream Computation in the Cloud. *ICDE*, 2015.

[25] Zaharia, M., Chowdhury, M., Das, T., Dave, A., Ma, J., McCauley, M., Franklin, M. J., Shenker, S., Stoica, I. Resilient Distributed Datasets: A Fault-Tolerant Abstraction for In-Memory Cluster Computing. *NSDI*, 2012.

[26] Zaharia, M., Das, T., Li, H., Shenker, S., Stoica, I. Discretize Streams: an Efficient and Fault-Tolerant Model for Stream Processing on Large Clusters. *SOSP*, 2013.

[27] Zhu, Y., Rundensteiner, E., Heineman, G. T. Dynamic Plan Migration for Continuous Queries over Data Streams. *SIGMOD*, 2014.

# 9 Appendices

## 9.1 Appendix A: Example of CH

Consider 12 tasks, among which 10 tasks have workload $w$, and the remaining 2 have workload $2w$. Given 2 nodes, a good assignment achieving load balancing allocates each node 5 tasks with workload $w$ and 1 task with workload $2w$. CH, however, only aims at that each node is assigned 6 tasks; hence, it is possible that one node may be allocated both of the heavy tasks. Further, consider that we add one more node. The best re-assignment is to move the 2 heavy tasks to the new node. CH, however, expects to move 4 tasks to the new node to balance the number of tasks on each node, leading to a higher migration cost.

## 9.2 Appendix B: Proofs

**Lemma 4.1:** Given a sub-problem $P=\langle[\alpha, \beta), [\gamma, \delta), n_P\rangle$. Let $maxgain(P)$ be the total gain obtained by the optimal solution of $P$. We have:

$$maxgain(P) = max_{x,y,n_l}\big(maxgain(P_1) + maxgain(P_2)\big)$$

where $P_1=\langle[\alpha, x), [\gamma, y+1), n_l\rangle$, $P_2=\langle[x, \beta), [y+1, \delta), n_P-n_l\rangle$, $\alpha<x<\beta-1$, $\gamma\leq y<\delta-2$, $1\leq n_l<n_P$.

**Proof:** It suffices to prove that there exists a particular combination of $x$, $y$ and $n_l$ such that $maxgain(P)=maxgain(P_1)+ maxgain(P_2)$. Assuming that we already know the optimal solution of $P$, denoted as $P^*$, we choose such a combination as follows. First, we select $y$ so that node $N_y$ is any node whose gain in $P^*$ is non-zero, meaning that $I_y\cap I'_y\neq\emptyset$, where $I_y$ and $I'_y$ (a part of $P^*$) are the task intervals of $N_y$ before and after migration, respectively. Then, we set $x$ to $I'_y.ub$. To determine $n_l$, we sort all task intervals in $P^*$ by their boundaries, and set $n_l$ to be number of intervals from the first up to $I'_y$ (inclusive).

Next we prove the lemma by contradiction. It is straightforward to prove that $maxgain(P)$ cannot be smaller than $maxgain(P_1)$ $+maxgain(P_2)$, since we can simply combine the optimal solutions of $P_1$ and $P_2$ to form a valid solution of $P$ with a total gain of $maxgain(P_1)+maxgain(P_2)$. Now suppose that $maxgain(P)> maxgain(P_1)+maxgain(P_2)$. Then, the additional gain of $P^*$ compared to the optimal solutions of $P_1$ and $P_2$ can only come from assigning a task in $P_1$ to a node in $P_2$, or vice versa. Without loss of generality, we assume the former case: specifically, in $P^*$, a task $T_z\in P_1$ ($\alpha\leq z<x$) is assigned to a node $N_w\in P_2$ ($y<w<\beta$), which obtains non-zero gain, i.e., $z\in I_w\cap I'_w$. Clearly, $z\notin I'_y$, as all tasks on $I'_y$ are assigned to node $N_y$. Because $I'_y.ub=x$, we have $z< I'_y.lb$. On the other hand, based on the assumption that nodes are ordered by their task intervals before the migration, we have $I_y.ub\leq I_w.lb$. Meanwhile, since $I_y\cap I'_y\neq\emptyset$, we have $z<I'_y.lb<I_y.ub<I_w.lb$, which contradicts with the assumption that $z\in I_w\cap I'_w$. □

**Lemma 4.2:** Given sub-problem $P$, there exists a combination of $x$, $y$, $n_l$ such that (i) $maxgain(P)=maxgain(P_1)+maxgain(P_2)$ and (ii) $N_y$ is the only node in $P_1$ with a non-zero gain in the optimal solution of $P_1$. The definitions of $P_1$, $P_2$ and the ranges of $x$, $y$, $n_l$ are the same as in Lemma 4.1.

**Proof:** (By contradiction) Let $x$, $y$, $n_l$ be the combination has the smallest $x$ among all combinations that satisfy condition (i). According to Lemma 4.1, such a combination exists. Suppose that in the optimal solution $P^*_1$ of $P_1$, there is another node $N_{y'}$ with non-zero gain, i.e., $I_{y'}\cap I'_{y'}\neq\emptyset$, where $I_{y'}$ and $I'_{y'}$ are the task intervals of $N_{y'}$ before and after migration, respectively. Let $x'= I'_{y'}.ub<x$, $n'_l$ be number of intervals in $P^*_1$ from the first up to $I'_y$ (inclusive), and $P'_1$, $P'_2$ be sub-problems defined using $x'$, $y'$, $n'_l$. Then, following the proof of Lemma 4.1, we have $maxgain(P)=maxgain(P'_1)$ $+maxgain(P'_2)$, which contradicts with the assumption that the combination of $x$, $y$, $z$ minimizes $x$ among all combinations satisfying condition (i). □

**Lemma 4.3:** Given sub-problem $P$ and a combination of $x$, $y$, $n_l$ that satisfies both conditions in Lemma 4.2. (i) The task interval $I_y$ of node $N_y$ before migration satisfies $I_y.lb<x$; (ii) if $I_{y+1}.ub<x$, then $maxgain(P_2)=maxgain(P'_2=\langle[x, \beta), [y+2, \delta), n_P-n_l\rangle)$.

**Proof:** Since node $N_y$ has a non-zero gain, we have $I_y\cap I'_y\neq\emptyset$, hence $I_y.lb\leq I'_y.ub$. According to Lemma 4.2, $I'_y.ub=x$, which proves (i). Regarding (ii), when $I_{y+1}.ub<x$, $I_{y+1}$ cannot possibly contain any task in $P_2$, meaning $N_{y+1}$ always has zero-gain. Hence, removing $N_{y+1}$ from $P_2$ does not affect the latter's maximum gain. □

**Lemma 4.4:** Given sub-problem $P$, there exists $x$ and $y$ such that the combination $x$, $y$, $n_{min}$ satisfies both conditions in Lemma 4.2, where $n_{min}$ is the minimum number of nodes for $P_1$ to have a feasible solution that satisfies load balancing given $x$ and $y$.

**Proof:** The maximum gain of $P_1$ computed by algorithm Solve_$P_1$ is independent of $n_l$. Meanwhile, increasing the number of target partitions for $P_2$ cannot possibly decrease its maximum gain, since we can always set a task interval to empty. Therefore, given any combination of $x$, $y$, $n_l>n_{min}$ satisfying the conditions of Lemma 4.2, we can obtain the same $maxgain(P_1)$ and $maxgain(P_2)$ with $x$, $y$, $n_{min}$, which also satisfies the conditions in Lemma 4.2. □

**Lemma 4.5:** Given sub-problem $P$, suppose that the combination $x$, $y$, $n_l$ satisfies both conditions in Lemma 4.2. Then, either of the following two conditions holds (i) $x-1\in I_y$, (ii) there does not exist another node $N_z$ in $P$ that leads to a higher gain for $P_1$, while satisfying $x\notin I_z$.

**Proof:** (by contradiction) Suppose that there exists a combination of $x$, $y$, $n_l$ that satisfies the conditions in Lemma 4.2, and at the same time violates both conditions

(i) and (ii) above. Let $I'_z$ be the task interval assigned to $N_z$ in the optimal solution $P^*$ of $P$. If the gain of $N_z$ is non-zero, i.e., $I_z \cap I'_z \neq \emptyset$, then, according to the second condition of Lemma 4.2, $N_z$ must belong to sub-problem $P_2$, hence $I'_z.ub \geq x$. Meanwhile, because $N_z$ leads to a higher gain for $P_1$ than $N_y$, $I_z$ must overlaps the task range of $P_1$, i.e., $[\alpha, x)$. Since $I_z \cap I'_z \neq \emptyset$, we have $x \in I_z$, leading to a contradiction. Therefore, the gain of $N_z$ must be zero, i.e., $I_z \cap I'_z = \emptyset$.

Let $I'_y$ be task interval assigned to $N_y$ in $P^*$. Now, consider reassigning $I'_y$ to $N_z$, and $I'_z$ to $N_y$. Since $N_z$ has a higher gain with $I'_y$ than $N_y$, this swap leads to a higher over all gain for $P$, which contradicts with the fact that $P^*$ is the optimal solution for $P$. □